\documentclass[8pt]{article}
\usepackage{amsmath,amsfonts}
\usepackage{graphicx}
\usepackage{hyperref}
\usepackage{verbatim}  % to use the comment environment

\newcommand{\be}{\begin{equation}}
\newcommand{\ee}{\end{equation}}
\newcommand{\matrice}{\begin{pmatrix}}
\newcommand{\ematrice}{\end{pmatrix}}

\newcommand{\bea}{\begin{eqnarray}}
\newcommand{\eea}{\end{eqnarray}}

   \let\g=\gamma  \let\d=\delta
        
              \let\r=\rho  
      \let\c=\chi
    
\let\G=\Gamma \let\D=\Delta   \let\L=\Lambda 
         
\let\Om=\Omega  \let\eps=\epsilon

\let\==\equiv
\let\io=\infty

\let\p=\partial

\def\to{\rightarrow}
\def\la{\left\langle}
\def\ra{\right\rangle}

\def\Tr{{\rm Tr}\,}
\def\la{\langle}
\def\ra{\rangle}

\def\Wsl{\,\raise.15ex\hbox{/}\mkern-13mu W}
\def\Dsl{\,\raise.15ex\hbox{/}\mkern-13mu D}
\def\Hsl{\,\raise.15ex\hbox{/}\mkern-12.5mu H}
\def\Lsl{\,\raise.15ex\hbox{/}\mkern-13mu L}
\def\hsl{\raise.15ex\hbox{/}\kern-.57em h}
\def\omsl{\raise.15ex\hbox{/}\kern-.5em\omega}
\def\intsl{\raise.15ex\hbox{/}\kern-.93em\int}

 at10pt

\begin{document}

\bigskip

\begin{center}
 {\LARGE\bfseries  Imposing causality on a matrix model}
\\[10mm]
Dario Benedetti, Joe Henson\\[3mm]
{\small\slshape
Perimeter Institute for Theoretical Physics \\
31 Caroline St.\ N, N2L 2Y5, Waterloo ON, Canada \\
%{\upshape\ttfamily dariobene@gmail.com} } \\[3mm]
}

\end{center}
\vspace{5mm}

\hrule\bigskip

\centerline{\bfseries Abstract} \medskip
\noindent
We introduce a new matrix model that describes Causal Dynamical Triangulations (CDT) in two dimensions.  In order to do so, we introduce a new, simpler definition of 2D CDT and show it to be equivalent to the old one.  The model makes use of ideas from dually weighted matrix models, combined with multi-matrix models, and can be studied by the method of character expansion.
\bigskip
\hrule\bigskip

\vspace{3cm}

%%%%%%%%%%%%%%%%%%%%%%%%%%%%%%%%%%%%%

\section{Introduction}

It has long been hoped that a theory of quantum gravity could be defined as the continuum limit of a discretised gravitational path integral.  Using the \textit{Dynamical Triangulations} (DTs) discretisation, the histories to be summed over are ``triangulations'' built by gluing together equilateral simplices in every possible configuration, weighted appropriately for gravity \cite{Ambjorn:book}.  With this simple definition, many results of physical interest can be obtained by analytical means.  In 2D, matrix model techniques \cite{DiFrancesco:2004qj} have been particularly useful in this regard, allowing the solution of many models of 2D quantum geometry and matter.

In 4D the original DT model does not seem to give physically acceptable results.  However, adding a physically motivated restriction on the space of configurations leads to the better-behaved Causal Dynamical Triangulation (CDT) models \cite{Ambjorn:2005qt}.  Here, all the observables that have been calculated are consistent with the emergence of an extended 4D geometry at large scales \cite{Ambjorn:2007jv,Ambjorn:2005db}.  The ability to calculate genuine generally covariant observables from a theory of quantum gravity, with good results, is a unique and important achievement for this approach.

However, not all of the analytical tools that were so useful in the case of DTs have been successfully carried over to the case of CDTs.  In the new model, the space of configurations is limited by demanding that: (\textit{a}) there exists a global time foliation; (\textit{b}) that spatial topology change is disallowed.  At first glance, these ``causal'' conditions seem to be of an irreducibly global nature.  Many of the analytical techniques used for DTs are not easily compatible with the imposition of such a global condition.
Other techniques have found a role in 2D CDTs, which indeed has been solved, in the pure gravity case, by several different means \cite{Ambjorn:1998xu,DiFrancesco:1999em}. Unfortunately none of these methods has proved useful in the presence of matter (where the only analytical results come from high-temperature expansion \cite{Benedetti:2006rv}), or in higher dimensions, with the only exception of the heap of pieces representation \cite{Benedetti:2007pp} of a reduced model of 3D CDT \cite{Dittrich:2005sy}.

In the original 2D DT model, the partition function can be re-expressed as a matrix integral, which in many cases can be explicitly computed.  The techniques have been generalised in many ways (\textit{e.g.} adding matter, different face shapes, and non-standard weights).  However, it has until now been thought impossible to impose the CDT conditions directly on a matrix model, and applications of matrix models to CDTs have followed other ideas \cite{Ambjorn:2008gk,Ambjorn:2008ta,Ambjorn:2008jf}.  Restrictions on the triangulations generated by a matrix model typically depend only on the properties of local structures such as vertices and faces.  It is not obvious how restrictions of this type could be used to introduce global foliations and ban topology change.

In this paper, we exhibit a novel matrix model that imposes the causal conditions, and generates CDT configurations.  As in the original DT case, the correspondence between the matrix model and CDTs applies directly at the discrete level.  The fact that this is possible is interesting in itself, opening up new possibilities for the study of matrix models and CDTs in general.   Using this matrix model, we can address problems in the CDT model, such as finding the critical point.  The results of these calculations will be presented elsewhere.

The observation that makes a matrix model formulation of CDTs possible is that global restrictions, such as topological restrictions, can sometimes be brought about as consequences of local restrictions, by what might be called ``rigidity''.  For instance, if a finite tessellation is restricted to have only four-sided faces, and only 4-valent vertices, it is easy to see that the only allowed topology is the torus, since the tesselation must be a regular lattice. This rigidity has been noted in the context of matrix models before, where ``dually weighted models'' have been developed that allow non-trivial weights that depend on both the valency of vertices and the number of sides of faces \cite{DiFrancesco:1992cn,Kazakov:1995ae,Kazakov:1998qw}.  The CDT restriction can be expressed in a similar way, as a ``partial rigidity''.  In a 2D CDT, each vertex in the triangulation is incident on exactly two spacelike edges, that connect it to its neighbours in its 1D spatial slice.  This property, along with the spacelike/timelike colouring of the edges bounding the triangles, completely characterises 2D CDTs, as we prove below in section \ref{s:newdef}.

Expressed in this new way, the CDT condition can be imposed on a matrix model (see section \ref{s:matrix}).  The model generalises the techniques of dually weighted matrix models to impose the first restriction, and it is a 2-matrix model so that it incorporates the spacelike/timelike labeling.  We envisage a new application of the character expansion methods mentioned above as a possible way to solve the matrix model; some steps towards the solution are presented in section \ref{s:reduction}.

%%%%%%%%%%%%%%%%%%%%%%%%%%%%%%%%%%%%%

\section{A new way of defining 2D CDTs}
\label{s:newdef}

The following definition of 2D CDTs follows from the standard one \cite{Ambjorn:1998xu} by replacing the terminology of triangulations with that of the dual fatgraph.  Vertices in the triangulation are dual to faces bounded by loops (or ``boundary-components'') in the fatgraph as usual; now, spacelike edges in the triangulation are dual to timelike edges in the dual and \textit{vice-versa}.  Below we will use the colour labels $A$ and $B$ for spacelike and timelike (the same names that we later give the corresponding matrices in the matrix model).  In a CDT, triangles, which had two timelike edges and one spacelike edge, become ``$AAB$ vertices'' incident on two spacelike and one timelike edge in the dual.

A CDT fatgraph is a finite connected fatgraph obeying the following conditions. % (see fig. \ref{f:CDT}).  
\textit{(1)} All vertices are $AAB$.  \textit{(2)} There are disjoint sets of faces, which we call ``strips'', each of which is connected, and such that its members are glued on their timelike edges only to other members of the set (these are dual to cycles of spacelike edges in the triangulation).  Each face in the strips in the dual fatgraph has exactly two timelike edges glued to the adjacent faces in the same strip.  \textit{(3)} It is possible to label these strips $s_0,s_1,...,s_{T-1}$, such that the following holds: each face $l_i$ (with labels natural to the cyclic order of the strip) on strip $s_t$ is glued by spacelike edges to faces on strip $s_{t+1}$ (this addition is mod $T$ for toroidial topology, while the rule is ignored for strip $s_{T-1}$ for a cylinder).  These gluings exhaust all of the spacelike edges (apart from the initial and final strips in the case of a spherical CDT, which are then glued to an initial or final face bounded entirely by spacelike edges)\footnote{In this definition, there is nothing to stop self-gluings and double-gluings between vertices, so a spatial slice may have length 2 or 1.  Also we have included an initial and final face, dual to an initial or final vertex rather than a boundary, so that these CDTs are spherical in topology rather than cylindrical.}.  The original description of 2D CDTs in \cite{Ambjorn:1998xu} mentions other properties, but they follow easily from those given here.

%\begin{figure*}
%\centering \resizebox{4.2in}{1.7in}{\includegraphics{cdt-vs-dual.eps}}
%\caption{\small{
%Example of a triangulated piece of spacetime in the CDT model
%(left), and the corresponding dual graph (right). We consider here the case where left and right boundary of the triangulation are identified so to form a cylinder.
%In the dual this means that the horizontal lines do actually form cycles.
%\label{f:CDT}
%}}
%\end{figure*}

This formulation is not suited to the application of matrix model techniques.  For this reason, we introduce an alternative definition: a CDT fatgraph is a finite connected fatgraph with only $AAB$ vertices, and such that every face has either 2 or 0 timelike edges.  Below, we will show that the new definition is equivalent to the original CDT definition.

\begin{figure*}
(a)\hspace{.5cm}
\centering \resizebox{4.2in}{1.0in}{\includegraphics{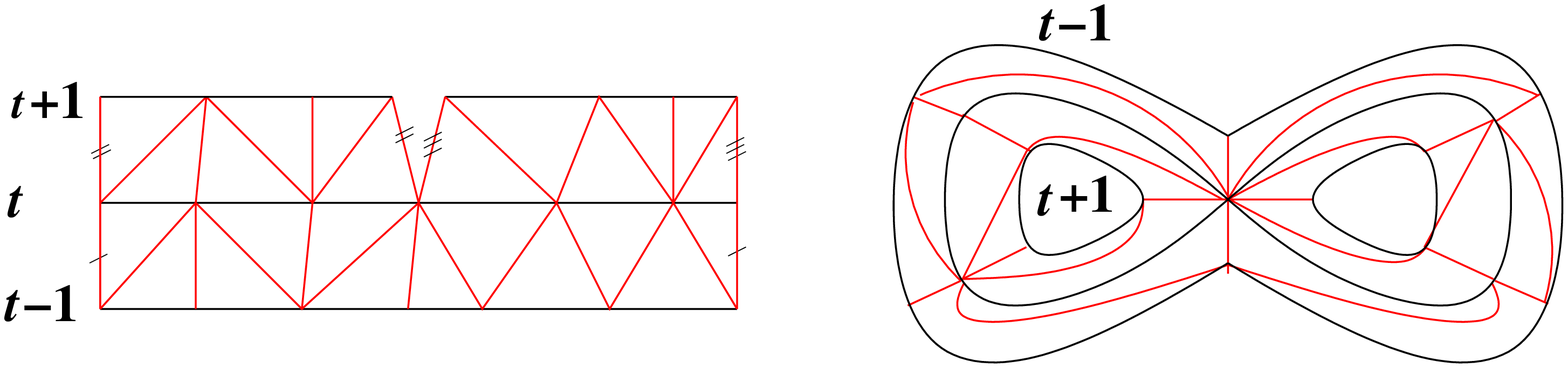}}\\
\vspace{.5cm}
(b)\hspace{.5cm}
\centering \resizebox{3.5in}{1.0in}{\includegraphics{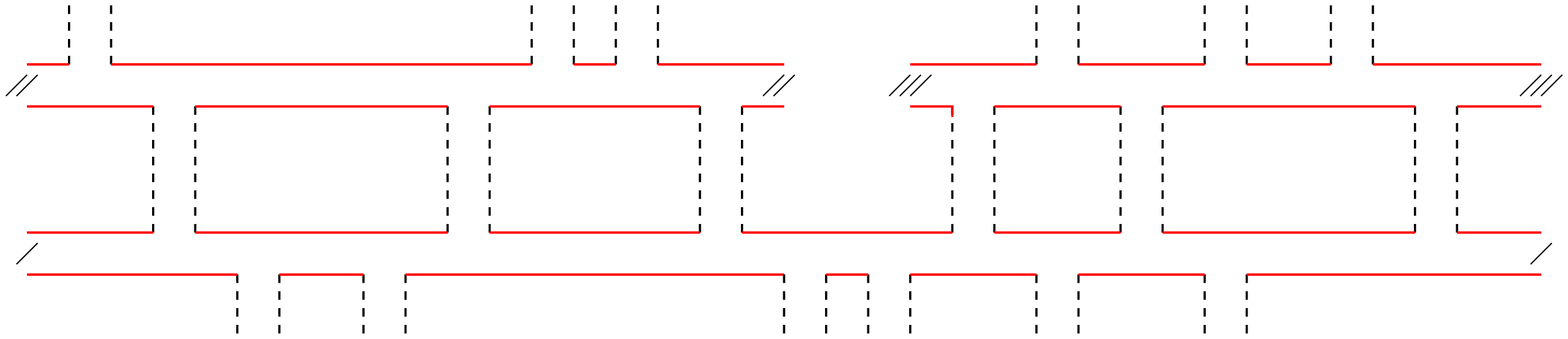}}
\caption{\small{
{\bf (a):} A triangulation with time slices, and with triangles with 2 timelike edges and one spacelike edge, but with spatial topology change.  Timelike edges are shown in red, and the slashes indicate identified edges.  The same triangulation is shown twice to illustrate both the time slices and the fact that, in time slice $t$, there is a branching vertex incident on more than 2 spacelike edges.
{\bf (b):} The dual fatgraph of the same triangulation.
\label{f:morse}
}}
\end{figure*}

One of the crucial features of the definition of 2D CDTs is the lack of spatial topology change: each time slice consists of one connected strip.  The new definition reveals that a more local restriction, that is, a restriction on the colourings of vertices and loops in the fatgraph, can be used to enforce this topological property.  In order for there to be topology change, one of two things would have to be allowed.  If all vertices are to remain $AAB$, the only way is to allow ``discrete Morse points'' (or ``branching points'') in the triangulation \cite{Ambjorn:1998xu}.  As shown in figure \ref{f:morse}, such a branching point has more than two spacelike edges incident on it, leading to a face in the dual with more than two timelike edges.  Another way of introducing topology change would be to add branching points in the dual, so that one strip would be glued to more than one strip in the same time direction.  This cannot happen without there being a vertex incident on more than 2 spacelike edges; the fact that all vertices are $AAB$ is therefore also crucial.  The comments below indicate how this CDT gluing between strips follows from the new definition.

It can immediately be seen that a fatgraph satisfying the original CDT definition also satisfies the new CDT definition given above.  The converse is not so obvious; to show this, we must take a fatgraph $\Gamma$ conforming to the new definition and show that it obeys each part of the (lengthier) original definition in turn.  Part \textit{(1)} of the definition is also a part of the new definition. As for part \textit{(2)}, in such a fatgraph $\Gamma$ the set of faces containing two timelike edges can be partitioned based on which faces glue on timelike edges (this is dual to partitioning vertices by connectivity on spacelike edges), which is a partitioning into strips as described above.  This leaves part \textit{(3)} of the original definition, which is less trivial.

Let us consider the topology of a strip $s$ in $\Gamma$, considered as a set of faces joined only by the gluings of their timelike edges.  The boundary is then the set of all spacelike edges in the strip.  Because every fatgraph corresponds to an orientable surface (see \textit{e.g.} \cite{David:1984tx,DiFrancesco:2004qj}), and each face is connected to only two others in the strip, the strip has the topology of a standard strip (as opposed to, say, a M\"obius strip).  The boundary of such a strip is composed of two cycles.  This observation allows us to show that the fatgraph does obey the third part of the original CDT definition.

First let us treat the case in which $\Gamma$ contains no completely spacelike faces (\i.e. faces containing no timelike edges).  In this case, all spacelike edges are in the boundaries of strips.  Each strip is glued to other strips (or to itself) by its spacelike edges, which, as previously shown, are partitioned into two spacelike cycles per strip.  If one such cycle was glued to spacelike edges in more than one other such cycle, three spacelike edges would somewhere meet at a vertex, contradicting our assumptions.  Thus, each strip is glued to only two others.  The collection of strips themselves, with the relation between them of gluing, must form a cycle (they cannot form multiple cycles or otherwise the graph would be disconnected, which is disallowed by assumption).  We can now choose a labelling of strips $s_0,s_1,...,s_{T-1}$ consistent with this cyclic order on the strips.  These gluings exhaust all of the spacelike edges.  This completes the proof that these fatgraphs obey the original CDT definition -- more specifically, the definition of toroidial CDTs.

There remains the case in which $\Gamma$ contains some completely spacelike faces.  The boundary loop of a completely spacelike face is a spacelike cycle.  Again, this must be glued to exactly one other spacelike cycle.  It is easy to see that only 2 such faces with no timelike edges can be included in the fatgraph if it is to be connected.  Hence we can repeat most of the argument from the former case.  In this case, it is possible to label the strips $s_0,s_1,...,s_{T-1}$, such that the following holds: faces in strip $s_t$ are glued by spacelike edges to faces on strip $s_{t+1}$, apart from when $t=T-1$.  These gluings exhaust all of the spacelike edges, except those from strips $s_0$ and $s_{t-1}$, which also glue to two completely spacelike faces.  Thus, these graphs fulfil the definition of spherical CDTs given above.

%%%%%%%%%%%%%%%%%%%%%%%%%%%%%%%%%%%%%

\section{The matrix model}
\label{s:matrix}

We are now going to define a two-matrix model that generates, via its perturbative expansion, the fatgraphs discussed in the previous section, and
whose partition function $Z$ will be related to that of the CDT model in the usual way, $i.e.$
\be \label{z-cdt}
Z_{CDT}=\lim_{N\to\io}\ln Z \ .
\ee
The two-matrix model is defined by its partition function
\be \label{zeta}
Z=\int dAdB \, e^{-N \Tr [\frac{1}{2}A^2+\frac{1}{2}(C^{-1}B)^2-g A^2 B]}\  ,
\ee
with $C$ an external matrix satisfying (in the large $N$ limit) the condition
\be \label{c}
\Tr(C^m) = N \d_{2,m}  \hspace{1cm}\text{for $m\geq 1$.}
\ee
It is not hard to see that such model generates graphs for which in each face there are always two (or zero) $B$-lines, and as
explained in the previous section this is enough to
impose the ``CDT condition" on the dual triangulation.
From standard analysis we can read off the free ($g=0$) propagators of the matrix model:
\be
\la A_{ij}A_{kl}\ra_0 = \frac{1}{N} \d_{il}\d_{kj}\ ,
\ee
\be
\la B_{ij}B_{kl}\ra_0 = \frac{1}{N} C_{il}C_{kj}\ ,
\ee
\be
\la A_{ij}B_{kl}\ra_0 = 0\ ,
\ee
from which we find that in the expansion in Feynmann graphs of (\ref{zeta}) a face contributes a factor of $\Tr(C^m)$ with $m$ being the number of internal $B$-lines,
and hence we can control such number by imposing conditions on  $\Tr(C^m)$, as in (\ref{c}).
Note that this is a generalization of the standard case where, being $C=1$, a face contributes just a factor $N$. On the other hand and just as usual
each edge brings a factor $\frac{1}{N}$ and each vertex a factor $g N$. In both cases the factors $N$ gather to give for each graph $\G$ a global $N^{\c(\G)}$
where $\c(\G) = F-E+V = 2-2h$ is its Euler characteristic ($F$, $E$ and $V$ are respectively the number of faces, edges, vertices of the graph and $h$ the number of handles
of the surface on which it can be drawn), so that the expansion in $1/N$ is effectively a topological expansion, as first shown by 't Hooft \cite{'tHooft:1973jz}.

The model can be simplified by noting that we can perform the integration over $B$, because it is a simple Gaussian, and obtain (up to a multiplicative constant which we discard from now on in the expression of the partition function)
\be \label{zeta-1matrix}
Z=  \int dA e^{-N\Tr[\frac{1}{2}A^2-\frac{g^2}{2} (A^2 C)^2]}\ ,
\ee
which looks similar to the standard one-matrix model which describes two-dimensional Euclidean quantum gravity and which was originally solved in \cite{Brezin:1977sv}
(note that this kind of reduction from two matrices to one matrix was considered in \cite{Beirl:1996cd} for
the case $C=1$ showing that a model with Lorentzian triangles without further restrictions would be in the same universality class as the standard Euclidean model) but with the important modification in the vertex which turns out to be crucial.
One can understand the integration over the matrix $B$ as the gluing of triangles along their spacelike edge, which can be done only in one way, and which gives rise (if there are no boundaries) to a model of squares with only timelike edges.
The presence of the external matrix $C$ introduces an anisotropy for the squares, while the condition (\ref{c}) acts to rigidly transfer such
anisotropy from local to global level, $i.e.$ to the whole quadrangulation.

%%%%%%%%%%%%%%%%%%%%%%%%%%%%%%%%%%%%%

\subsection{Character expansion}
\label{s:reduction}

The standard reduction to eigenvalues is not directly available in our case because, as  $[A,C]\neq 0$,  the two matrices are not simultaneously diagonalizable, so we need some fancier
trick to solve the model, like the character expansion \cite{DiFrancesco:1992cn,Kazakov:1995ae,Kazakov:1998qw}.
Indeed in terms of reduction to eigenvalues our interaction term poses exactly the same problem as the $ABAB$ interaction of \cite{Kazakov:1998qw}\footnote{Just replace their $A$ with our $A^2$
and their $B$ with our $C$. Of course our is a completely different model, because of the interaction containing four powers of $A$ rather than two, because $C$ is an external matrix,
and because we have no other self-interaction term; but what we are interested in here is the similar $xyxy$ structure which we want to disentangle.}, $i.e.$ we have to perform the integral
\be
\int d \Om\ e^{N\frac{g^2}{2} \Tr(\Om \L^2 \Om^\dag C)^2}\ ,
\ee
where $\L$ is the diagonal matrix of eigenvalues of $A$. Unlike the case of the Itzykson-Zuber-Harish-Chandra integral (which differs from our problem for the absence of the overall square
in the exponent) there is no exact formula for this kind of integral.
Following \cite{Kazakov:1998qw} we can instead perform the character expansion
\be \label{ch-exp}
e^{N\frac{g^2}{2} \Tr(A^2 C)^2} \sim \sum_{\{h\}} \Big(N\frac{g^2}{2}\Big)^{\# h/2} \frac{\D(h^e)\D(h^o)}{\prod_i (\tfrac{h_i^e}{2})! (\tfrac{h_i^o-1}{2})!}
  \text{sgn} [ \prod_{i,j} (h^e_i-h^o_j)  ] \chi_{\{h\}}(A^2 C)\ ,
\ee
where the sum is over all integers $h_i$ such that $h_N>h_{N-1}>...>h_1\geq0$ and which have to be in equal number even ($h^e$) and odd ($h^o$),
$\chi_{\{h\}}(x)$ is the character of the group element $x$ (the group is $GL(N)$ in our case) in the representation labelled by the set of integers $\{h\}$, $\# h=\sum h_i -\tfrac{1}{2}N(N-1)$ is the total number of boxes in the corresponding Young tableau, and $\D(x)$ is the Vandermonde determinant.

Now we can perform the integration over the unitary group using the character orthogonality relation
$\int d \Om \, \chi_{\{h\}}(\Om  \L_A^2 \Om^\dag \L_C)= \chi_{\{h\}}(A^2)\chi_{\{h\}}(C)/d_{\{h\}}$ (where $d_{\{h\}}$ is the
dimension of the representation given by $d_{\{h\}} = \D (h)/ \prod^{N-1}_{i=1} i!$), to obtain
\be \label{zeta-expanded}
Z \sim   \sum_{\{h\}} (N\frac{g^2}{2})^{\# h/2}  c_{\{h\}} \chi_{\{h\}}(C) I_{\{h\}}\ ,
\ee
where $c_{\{h\}}$ is the following coefficient ($[x]$ denotes the integer part),
\be \label{coeff-c}
c_{\{h\}} \equiv \frac{1}{\prod_i [ h_i/2] ! \prod_{i,j}(h_i^{e}-h_j^{o})} \ ,
\ee
and $I_{\{h\}}$ is the matrix integral
\be \label{ch-int}
I_{\{h\}} \equiv  \int dA\ \chi_{\{h\}}(A^2) e^{-\frac{N}{2}\Tr A^2}  \equiv  \la \chi_{\{h\}}(A^2) \ra_0\ ,
\ee
which we have rewritten in terms of the full matrix for esthetic reasons (the angular integration is now just an overall constant, and hence we have the freedom
to move it in and out of the partition function).
Using the formula
\be \label{ch-c}
\chi_{\{h\}}(C) = c \big(\frac{N}{2}\big)^{\tfrac{1}{2}\sum_i h_i} \frac{\D(h^e)\D(h^o)}{\prod_i (\tfrac{h_i^e}{2})! (\tfrac{h_i^o-1}{2})!}\text{sgn} [ \prod_{i,j} (h^e_i-h^o_j)  ]\ ,
\ee
originally given in \cite{Kazakov:1995ae}, for the character of a matrix satisfying the condition  (\ref{c}),
we can rewrite $Z$ as
\be \label{our-z}
Z \sim  \sum_{\{h\}} (N\frac{g}{2})^{\# h}  c^2_{\{h\}} \D(h) I_{\{h\}}\ .
\ee
We have now succeeded in reducing the original integral in $O(N^2)$ variables to a problem in $O(N)$ variables, and this is a promising step for its solution.

%%%%%%%%%%%%%%%%%%%%%%%%%%%%%%%%%%%%%%%%%%%%%%%%%%%%%%%%%%%%%%%%%%%

%\subsection{The critical point}
%\label{s:cpoint}

Unfortunately the integral (\ref{ch-int}) is not of the Di Francesco-Itzykson type \cite{DiFrancesco:1992cn}, and an exact formula is not known.
Another possible way to tackle the evaluation of the partition function (\ref{zeta-expanded}) could be a double saddle point method along the lines of \cite{Kazakov:1998qw},
but again the quadratic dependence on $A$ in the character makes the existing results difficult to use, and we could not obtain sensible saddle point
equations.

We could of course simply rely on the equivalence of the model to usual CDT, as proven in Sec.~\ref{s:newdef}, and hence conclude that the model has already been solved. Although that would be a completely legitimate attitude, we find it instructive to try and take some further steps within the character expansion method,
especially in view of the fact that, if successful, this method might be extended to solve other CDT models that have not been solved by other means.

In order to proceed let us then look for a moment at the partition function $Z_1$ for the usual matrix model with quartic vertex, which is obtained by substituting $C=1$ in (\ref{zeta-1matrix}), and which was first solved in  \cite{Brezin:1977sv}.
If in (\ref{zeta-expanded}) we substitute $C=1$, we find
\be \label{usual-z}
\begin{split}
Z_1  &\sim   \sum_{\{h\}} (N\frac{g^2}{2})^{\frac{\# h}{2}}  c_{\{h\}} d_{\{h\}} I_{\{h\}}\\
    &=   \sum_{\{h\}} (N\frac{g^2}{2})^{\frac{\# h}{2}} \frac{\D(h^e)\D(h^o)}{\prod_i \left[\frac{h_i}{2}\right]!} \la \chi_{\{h\}}(A^2) \ra_0\ ,
\end{split}
\ee
which of course in this expansion presents the same difficulty as our new model.
In order to find an expression for $\la \chi_{\{h\}}(A^2) \ra_0$ we compare (\ref{usual-z}) to the expansion
that was derived for the same model in \cite{Kazakov:1995ae},
\be \label{kazakov-exp}
Z_1 \sim \sum_{\{h^{(0)},h^{(1)},h^{(2)},h^{(3)}\}} \left( \frac{g^2}{2N} \right)^{\frac{\# h}{4}} \prod_{\eps=0}^3 \frac{\D^2(h^{(\eps)}) \prod_i h_i^{(\eps)}!!}{\prod_i\big(\frac{h_i^{(\eps)}-\eps}{4}\big)!}
   \prod_{i,j} (h_i^{(0)}-h_j^{(2)}) (h_i^{(1)}-h_j^{(3)})\ , %\prod_i (h_i^{(0)}-1)!!(h_i^{(2)}-1)!! h_i^{(1)}!! h_i^{(3)}!!
\ee
where the integers $h$ have factored into 4 groups of $N/4$ integers $h^{(\eps)}$ with $\eps=0,1,2,3$ denoting the congruence modulo 4.
Matching powers of $g$ among (\ref{usual-z}) and (\ref{kazakov-exp}), we can conjecture
\be \label{DFI-2}
\la \chi_{\{h\}}(A^2) \ra_0 \sim \frac{1}{N^{\# h}} \prod_{\eps=0}^3 \D^2(2 h^{(\eps)}) \prod_i (2 h_i^{(\eps)})!!\ ,
\ee
as the extension of the Di Francesco-Itzykson formula we were looking for. We stress that this is by no means a proof, but just a conjecture that we would like to test against the known solution of CDT.

%%%%%%%%%

Substituting (\ref{DFI-2}) into our partition function (\ref{our-z}) we reduce it to a pure sum over representations,
\be \label{our-z-2}
Z \sim \sum_{\{h^{(0)},h^{(1)},h^{(2)},h^{(3)}\}} \left( \frac{g}{2} \right)^{\frac{\# h}{2}} \prod_{\eps=0}^3 \frac{\D^2(h^{(\eps)}) \prod_i h_i^{(\eps)}!!}{\left(\prod_i\big(\frac{h_i^{(\eps)}-\eps}{4}\big)!\right)^2}
   \frac{\prod_{i,j} (h_i^{(0)}-h_j^{(2)}) (h_i^{(1)}-h_j^{(3)})}{\prod_{i,j} (h_i^e-h_j^o)}\ .
\ee
We can now proceed to try and solve this model via a saddle point method for the  shifted highest weights.

Similarly to what is usually done in the case of the eigenvalues, in the large $N$ limit it is useful to rescale $h_i\to h_i/N$,
introduce the density distribution
%\be
$\rho(h) = \tfrac{1}{N} \tfrac{\p i}{\p h}$,
%\ee
for which we assume (following \cite{Kazakov:1995ae,Kazakov:1998qw}) that
\be \label{rho-h}
\begin{split}
\r(h)=1 \ , & \ \ \ \ \text{for} \ \  0<h<h_1\ ,\\
0<\r(h)<1 \ , & \ \ \ \ \text{for}\ \   h_1<h<h_2\ ,
\end{split}
\ee
for some $h_1$ and $h_2$ to be determined (note that the indices 1 and 2 have nothing to do with the index $i$ which is omitted in the continuous notation).
This condition is simply the statement that in a typical Young tableau of $N$ rows there will be a lower fraction of rows which are empty.
Finally we define the resolvent function
\be \label{resolvent}
H(h) = \int_0^{h_2} d h' \frac{\r(h')}{h-h'} = \ln \frac{h}{h-h_1} + \int_{h_1}^{h_2} d h' \frac{\r(h')}{h-h'} = \ln \frac{h}{h-h_1}+ \tilde H(h)\ ,
\ee
which as usual satisfies
\be
\Hsl (h_i) = \frac{1}{N} \frac{\p}{\p h_i} \log \D(h)\ , \hspace{1cm} \text{for $h_i\in [0,h_2]$}\ ,
\ee
where we used the notation $\Hsl (h) = \tfrac{1}{2}(H(h+i\eps)+H(h-i\eps) $.
Assuming that, because of symmetry, the various sets of weights should have the same distribution,
we arrive at the following saddle point equation:
\be
\tilde \Hsl(h)= -2 \ln (2g) -\ln \frac{h}{h-h_1} \ ,  \hspace{1cm} \text{for $h\in [h_1,h_2]$}\ .
\ee
This equation constitutes a standard Hilbert problem \cite{Muskhelishvili:1992} whose solution is given by
\be
\tilde H(h) = \sqrt{(h-h_1)(h-h_2)} \oint_\g \frac{dz}{2\pi i} \frac{2\ln(2g)+\ln \frac{z}{z-h_1}}{(z-h)\sqrt{(z-h_1)(z-h_2)}}\ ,
\ee
with the contour $\g$ encircling counterclockwise the cut $[h_1,h_2]$.
The integral can be evaluated by inflating the contour and catching the pole at $z=h$ and the logarithmic cut on $[0,h_1]$.
Taking the signature of the square root to be $+$ to the right of the cut, and $-$ to the left (and $\pm i$ respectively above and below the cut),
and adding back the logarithmic part of the resolvent we find
\be
H(h) = \ln \left[ \frac{h_2-h_1}{4 g^2} \frac{h}{h (h_1+h_2)-2 h_1 h_2 +2\sqrt{h_1 h_2} \sqrt{ (h-h_1)(h-h_2)}}  \right]\ .
\ee
On this expression we have to impose the condition $H(h)\sim \tfrac{1}{h}$ for large $h$, which is a consequence of the normalization
of the density function $\rho(h)$, and this fixes the parameters $h_1$ and $h_2$ to be $h_1 = \frac{1-4 g^2}{1+4g^2}$ and $h_2 = \frac{1+4g^2}{1-4g^2}$.
As expected we find a critical point at $g_c=1/2$, which is exactly the critical point of CDT \cite{Ambjorn:1998xu}.
At the critical point $h_1=0$ and $h_2=+\io$, $i.e.$ the distribution has no saturated part and is excited all the way to $h=+\io$.

To turn the above derivation into a proper solution it would of course be fundamental to prove the conjecture behind formula (\ref{DFI-2}), but it is encouraging to see that this gives the right result for the critical point.

%%%%%%%%%%%%%%%%%%%%%%%%%%%%%%%%%%%%%%%%%%%%%%%%%%%%%%%%%%%%%%%%%%%

\section{Conclusion}

Matrix model techniques proved to be a powerful tool in the study of 2D gravity before the introduction of CDTs.  By combining ideas from multi-matrix models and dually weighted matrix models, we have been able to extend the technique to CDTs in a direct way.  This has been achieved using the idea of ``rigidity'': that constraints on local structures in a fatgraph can have global consequences.  As we have seen, a restriction on colourings of vertices and faces serves to impose the CDT condition with all its consequences (for example, the restriction on global topology).  It has been observed that the CDT model is intermediate between the unrestricted dynamical spacetime of the DT model, and the totally rigid regular lattice.  It is interesting to review this observation in the light of the above results.

The standard CDT model of the type discussed above rules out spatial topology change, and it is interesting to consider re-admitting topology change in a controlled way in the above model, as has been done in \cite{Ambjorn:2008ta}.  As mentioned in section \ref{s:newdef} this can be done by allowing vertices connecting to four spacelike edges (\textit{e.g.} by allowing $A^4$ vertices), or faces with four timelike edges (the dual of this is shown in figure \ref{f:morse}).  The latter case would mean altering (\ref{c}), setting $\Tr(C^m) = N \d_{2,m} + N c \d_{4,m}$ for $m\geq 1$, where $c$ is a new ``dual coupling'' constant.
Of course adding such extensions without control would bring us back to the universality class of the Euclidean DT, so we would instead have to scale correctly the new coupling in the continuum limit to control the topology change, in a similar fashion to what done in \cite{Ambjorn:2008ta}. It would be interesting to compare the results of this approach with those of \cite{Ambjorn:2008gk,Ambjorn:2008ta,Ambjorn:2008jf} and maybe establish a direct link.

The matrix model above will be studied further in a forthcoming paper, using generalisations of the character expansion methods of \cite{DiFrancesco:1992cn,Kazakov:1995ae,Kazakov:1998qw}. Hopefully, the power of matrix model techniques can then be applied to unsolved problems in 2D CDTs.  Outstanding among these is the addition of matter, and the derivation of interesting results like those found for the ``pure gravity'' case.  By comparison to previous work on DT matrix models, one could write down models for these cases without great difficulty.  For example, one way of putting the Ising model on the CDT matrix model would be to double the number of matrices, replacing $A$ with $A_+$ and $A_-$ (and perhaps also $B$ with $B_+$ and $B_-$) in an analogous way to the case of DTs.  Application to the Potts model and others would be similar.

The new perspective given here on CDTs is neither confined to 2D nor to CDT.  In three and four dimensions also, analytic techniques that were of use in DTs \cite{Ambjorn:1990ge} have previously been inapplicable to CDTs, due to the form of the causal condition. However, there are extensions of the claims of section \ref{s:newdef} that can be conjectured to hold in higher dimensions.  This leads to a reformulation of the causal condition.  It would be most interesting to determine if this reformulation can overcome the barriers to analytic progress in CDTs in three and four dimensions.  Investigations of this question by Daniele Oriti, Pedro Machado and the authors are currently underway.
Furthermore these new way of imposing causality on the graphs might be of interest for other approaches to quantum gravity such as Group Field Theory \cite{Oriti:2006se}.

%%%%%%%%%%%%%%%%%%%%%%%%%%%%%%%%%%%%

\section*{Acknowledgements}

We thank Daniele Oriti for having posed the question that triggered this project.
We would also like to thank the Institute for Theoretical Physics in Utrecht, for hospitality during completion
of this work.
Research at Perimeter Institute for Theoretical Physics is supported in part by the Government of Canada through NSERC and by the Province of Ontario through MRI.

%%%%%%%%%%%%%%%%%%%%%%%%%%%%%%%%%%%%%

%%%%%%%%%%%%%%%%%%%%%%%%%%%%%%%%%%%%

%%%%%%%%%%%%%%%%%%%%%%%%%%%%%%%%%%%%

\providecommand{\href}[2]{#2}\begingroup\raggedright
\endgroup

\end{document}